\documentclass[preprint,12pt]{elsarticle}

\usepackage{amssymb}
\usepackage{physics}
\usepackage{amsmath,amssymb}
\usepackage{graphicx}
\usepackage{mathrsfs}
\usepackage{subfigure}

\begin{document}
	
	\begin{frontmatter}

\title{The effect of vacancy induced localized states on thermoelectric properties of armchair bilayer phosphorene nanoribbons }

\author[label1]{S. Jalilvand}
\author[label1]{S. Sodagar}
\author[label2]{Z. Noorinejad}
\author[label3]{H. Karbaschi}
\author[label1]{M. Soltani}
\address[label1]{Faculty of Physics, University of Isfahan, Isfahan, Iran}
\address[label2]{Department of Physics, Islamic Azad University-Shareza Branch (IAUSH), Shahreza, Iran}
\address[label3]{Mahallat Institute of Higher Education, Mahallat 37811-51958, Iran}

\ead{mo.soltani@sci.ui.ac.ir}

\begin{abstract}
We consider an armchair bilayer phosphorene that is connected to two hot and cold leads from both sides and study the thermoelectric properties of such a system with periodic vacancies at the middle of nanoribbon and in the armchair direction. For this purpose, we first analytically show that by creating a vacancy, a localized state is generated around it. Then we demonstrate that in the presence periodic vacancies, a new energy band will be formed in the energy bandstructure, and by changing the distance between the vacancies, the width of the transmission channel and finally the electric power and thermoelectric efficiency can be tuned.
\end{abstract}


\end{frontmatter}

\section{Introduction}
\label{sec.intro}

	Due to the limited energy resources and the rapid increase in energy consumption, access to reliable and renewable sources of energy is becoming more and more critical.
The generation of electric current from the heat gradient is a process that occurs in thermoelectric generators (TEGs) and can be proposed as a solution to the energy shortage crisis.

The most fascinating feature of TEGs is that by employing waste heat, which is a free source of energy, they generate clean and eco-friendly energy. TEGs have no moving parts making them reliable, maintenance-free, and silent. Although the advantages of TEGs have been known to researchers for years, their use was very limited due to their low performance.

The theory of charge carrier confinement predicts that low-dimensional materials have unique properties that make them very different from bulk materials\cite{Hicks93a, Hicks93b}. Researchers believe that some of these properties can be used to improve thermoelectric properties \cite{Balandin2003, Murphy08, Esposito09, Leijnse10, Karbaschi2016, ZHANG2016, Hung2016, Karbaschi2020, Rezaei2021}.

In low-dimensional structures, novel quantum phenomena occur, some of which can improve the electrical power of thermoelectric systems.
Also, in low-dimensional structures, the number of system boundaries is much higher than in bulk materials, which increases phonon scattering and ultimately decreases thermal conductivity, which is favorable for thermoelectric applications.

Although it was thought that it is not possible to achieve two-dimensional structures, the discovery of graphene as the first two-dimensional material, demonstrated that it was possible to obtain stable sheets of atomically thin crystals with outstanding properties. 
The discovery of other two-dimensional materials which show variety of unique properties opened new attractive opportunities in the field of material science and condensed matter physics. 

Among the two-dimensional materials, the few-layers phosphorene has attracted much attention due to its special mechanical, electronic and optical properties \cite{Amini_2019,Sodagar_2023}. Similar to graphene, phosphorene nanoribbons have two basic edge structures, namely armchair and zigzag, but unlike graphene, phosphorene is a large and tunable energy bandgap semiconductor. 
In previous researches, it has been observed that phosphorene nanoribbons are anisotropic materials that have different properties in two directions\cite{Carvalho2016}. Also, it has been showed an outstanding properties of zigzag phosphorene nanoribbons is the existence of edge states in the middle of the bandgap and the conduction window of such edge states can improve the thermoelectric properties\cite{article3, Ezawa2014}.
Such edge states are not observed in the phosphorene nanoribbons with armchair edge but the interesting thing is that by adding a vacancy to armchair phosphorene nanoribbons, a localized state is created around the vacancy with energy in the bandgap region. The presence of this localized state in the energy bandgap will act like a single energy level atom\cite{Zare,Kirali}. If we periodically add a set of vacancies in the nanoribbon, due to the overlap of the wave functions of these states, there will be an interaction between them. Therefore, a periodic set of vacancies will act as a one-dimensional tight-binding chain of single energy level atoms with an impurity band in the energy band structure. The width of the conduction window of these modes can be tuned by changing the distance between vacancies, which is an important issue in the design of thermoelectric systems at different temperatures. In this research, we have first introduced bilayer phosphorene nanoribbons, and then we have shown with analytical and numerical calculations that the existance of vacancies leads to appreance of localized states around the vacancy and the periodicity of such vacancies results in a creation of an impurity band.
The remainder of this paper is organized as follows.In the next section, the tight binding formalism is described and the existence of localized states is proved analytically. In section III the formalism of thermoelectric properties calculations is introduced. Finally, the paper ends with the results and conclusions

\section{Model and Slution}

\begin{figure}[h!]
	\includegraphics[height=1.\linewidth]{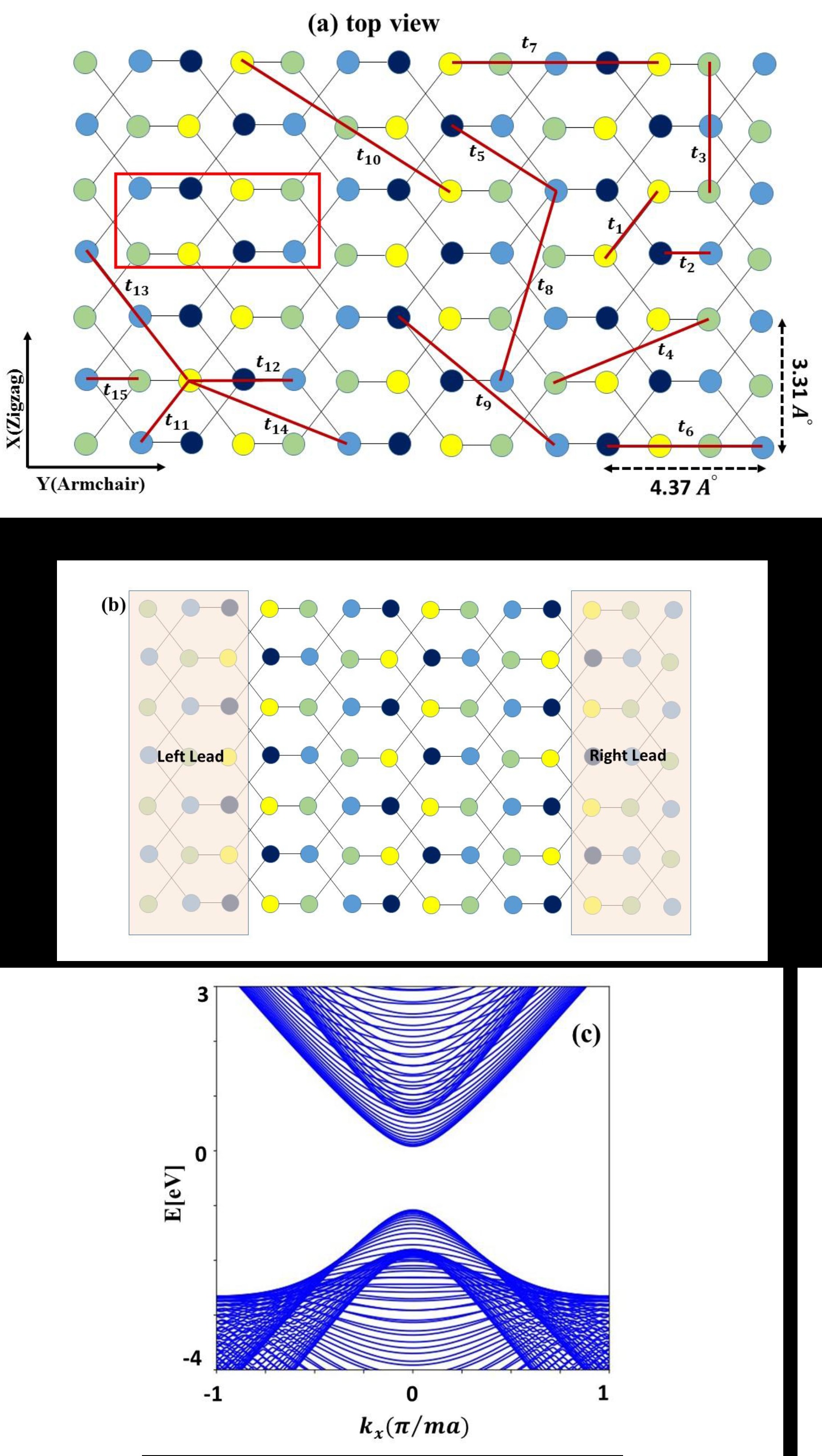}	
	\caption{\label{fig.bilayer}
		(a) five inter-layer and ten intra-layer hopping parameters in top view of bilayer phosphorene crystal structure. The red rectangle indicates the unit cell. (b) The shadow areas represent the source (left) and drain (right) semi-infinite leads. (c) Band structure of armchair bilayer phosphorene.
	}
\end{figure}

The crystal structure of a pristine armchair bilayer phosphorene used in our study is indicated in Fig\ref{fig.bilayer}. 

Bilayer phosphorene can be expressed as a two-dimensional material that consists of a non-planar puckered honeycomb lattice of phosphorous atoms with two inequivalent sublattices denoted by A and B.  In order to investigate the qualitative electronic properties and the energy band structure of bilayer phosphorene, we use a tight-binding approach in which the electrons are allowed to hop between neighbors sites with hopping energy, and the hamiltonian can be written as:
\begin{equation}
	\begin{aligned}
		H={}&\sum_{\langle i,j\rangle \langle i^\prime,j^\prime \rangle}t^{\parallel}_\mathbf{i}(\hat{c}^{\dag A_{1(2)}(B_{1(2)})}_{i,j}\hat{c}^{B_{1(2)}(A_{1(2)})}_{i^\prime,j^\prime}\\
		&+\sum_{\langle i,j \rangle \langle i^\prime,j^\prime \rangle}t^{\perp}_\mathbf{i}(\hat{c}^{\dag A_1(B_2)}_{i,j}\hat{c}^{B_2(A_1)}_{i^\prime,j^\prime}\\
		&+\sum_{\langle i,j \rangle}\epsilon_{i,j}(\hat{c}^{\dag A_{1(2)}(B_{1(2)})}_{i,j}\hat{c}^{A_{1(2)}(‌‌B_{1(2)})}_{i,j}),
	\end{aligned}
	\label{HAM}
\end{equation}
where the first and second terms describe the intra-layer and inter-layer Hamiltonian respectively.
In order to avoid confusion let us label each site by two indexes $(i, j)$  where $i$ denotes the row number along the $x$-direction and $j$ denotes the position of the site along the $y$-direction.
In Hamiltonian of Eq.~(\ref{HAM}), $t_\mathbf{i}$ describe hopping parameters between the $\langle i,j \rangle$ and $\langle i^\prime,j^\prime \rangle$ sites  which are shown in figure 1. $\hat{c}_{i,j}$($\hat{c}^{\dag}_{i,j}$) is the annihilation(creation) operator of electron at site$\langle i,j\rangle$ and $\epsilon_{i,j}$ is the on-site energy and there are set to zero for all lattice sites when no defect and external field exist in the lattice. Vacancies are an important part of defects that play a paramount role in the electronic properties of the phosphorene structures. The simulation of the vacancy in hamiltonian is modeled by setting the on-site energy equal to a very large value. Therefore the hopping energies between vacancy sites and the nearest neighbors disappear. Table 1 supports the fifteen relevant tight binding hopping parameters, ten intra-layer, and five inter-layer. 
\begin{table}[h!]
	\centering
	\caption{\footnotesize 10 intra-layer ($t^{\parallel}$) and 5 inter-layer ($t^{\perp}$) hopping parameters of bilayer phosphorene, shown in figure \ref{fig.bilayer}(a).}
	\label{tab.hopping}
	\begin{tabular}{|c|c|c|c|c|c|c|c|c|}
		\hline
		$\mathbf{i}$ & $t^{\parallel}_\mathbf{i}$ &$d_\mathbf{i}(\mathrm{\AA})$&$\mathbf{i}$ & $t^{\parallel}_\mathbf{i}$ &$d_\mathbf{i}(\mathrm{\AA})$&$\mathbf{i}$ & $t^{\perp}_\mathbf{i}$ &$d_\mathbf{i}(\mathrm{\AA})$\\
		\hline1&-1.486&2.22&6&0.186&4.23&11&0.524&3.60\\
		2&3.729&2.24&7&-0.063&4.37&12&0.180&3.81\\
		3&-0.252&3.31&8&0.101&5.18&13&-0.123&5.05\\
		4&-0.071&3.34&9&-0.042&5.37&14&-0.168&5.08\\
		5&0.019&3.47&10&0.073&5.49&15&0.005&5.44\\
		\hline
	\end{tabular}
\end{table}
The multilayer phosphorene nanoribbon has a highly anisotropic structure and cutting the phosphorene into the horizontal (y-direction) yields an armchair edge while cutting along the vertical (x-direction) yielding a zigzag edge.The presence of a zigzag edge in nanoribbon causes a quasi flat edge mode in the Fermi energy range. On the contrary, it is demonstrated that the armchair edge leads to the disappearance of edge modes in the energy difference between the conduction band minimum (CBM) and valance band maximum (VBM).The band structure of a pristine Armchair bilayer phosphorene used in our study is indicated in Fig\ref{fig.bilayer}. 
Besides, it is worth mentioning that inner vacancy can induce a localized state around the vacancy position. It should be noted that by placing the vacancy site on the A(B)-type sublattice, a wave function of localized state appears on the B(A)-tape sublattice. Also, by moving away from the vacancy position, the absolute value of wave function decreases.\par 
As we discussed in Ref(1), we divided Hamamiltonian into two parts: $\hat{H_{0}}$ and $\hat{H_{1}}$. It can be considered that $t^{\parallel}_1,t^{\parallel}_2,t^{\parallel}_6$ and $t^{\perp}_{11}$ hopping parameters are related to $\hat{H_{0}}$ and other hopping parameters are applied to $\hat{H_{1}}$ subsequently. As we discussed in(), the edge states have the following properties: $\hat{H_{0}}\ket{\psi_{edge}}=0$. In consequence, to obtain an impurity state around the vacancy site, we look for a wave function in which the following relation establish:
\begin{equation}
	\hat{H_{0}}\ket{\psi_{\mathcal{V}}}=0
\end{equation}
In the first step of article, for simplicity, we calculate the localized state near the single vacancy site analytically by using $t^{\parallel}_1,t^{\parallel}_2$ and $t^{\perp}_{11}$ hopping parameters. In the continuation of the investigation, the effect of other hopping parameters as perturbation terms calculate numerically, which shifts the energy value.
\begin{equation}
	E_\mathcal{V}=\bra{\psi_{\mathcal{V}}}\hat{H_1}\ket{\psi_{\mathcal{V}}}
\end{equation}
\subsection{Localized States Around Vacancy In Bulk Armchair Bilayer Phosphorene Nanoribbon} 
Bulk properties of materials such as electronic properties, specific heat, and other transport phenomena are directly proportional to the local density of states(LDOS). In order to discuss the local density of states(LDOS), one should start with Green's function. The retarded Green's function can be calculated from the Landauer-Buttiker formula as 
\begin{equation}
	G(\epsilon)=\frac{1}{\epsilon-H_{center}-\Sigma^{r}_L(\epsilon)-\Sigma^{a}_R(\epsilon)}
\end{equation}
where $H_{center}$, $\Sigma^{r}_L(\epsilon)$ and $\Sigma^{a}_R(\epsilon)$are the Hamiltonian of center region, the left and right retarded self-energy,respectively. To calculate the local density of states at a given site $i$ in a central conductor, we use the imaginary part of Green's function: 
\begin{equation}
	\rho_i(E)=\frac{-1}{\pi}Im[G(i,i)]
\end{equation}
where $\rho_i(E)$ represents the LDOS of system. By analyzing the local density of states, we will find that presence of a localized state beyond bulk states will cause a sharp peak in LDOS figures in energy space. For a localized state with zero energy we have:
\begin{equation}
	\rho_i(E)=\vert\braket{i|\psi}\vert^2\delta (E)
\end{equation}
In other words, the Dirac delta curve in the LDOS figures represents a localized state. Based on this description, it is expected that $H_0\ket{\psi_{\mathcal{V}}}=0$ and therefore, two conclusions can be drawn: first $\braket{B,i,j,k|\psi}=0$ and second $H_0\ket{\psi_{\mathcal{V}(A)}}=0$.

\begin{figure}[h!]
	\includegraphics[height=0.7\linewidth]{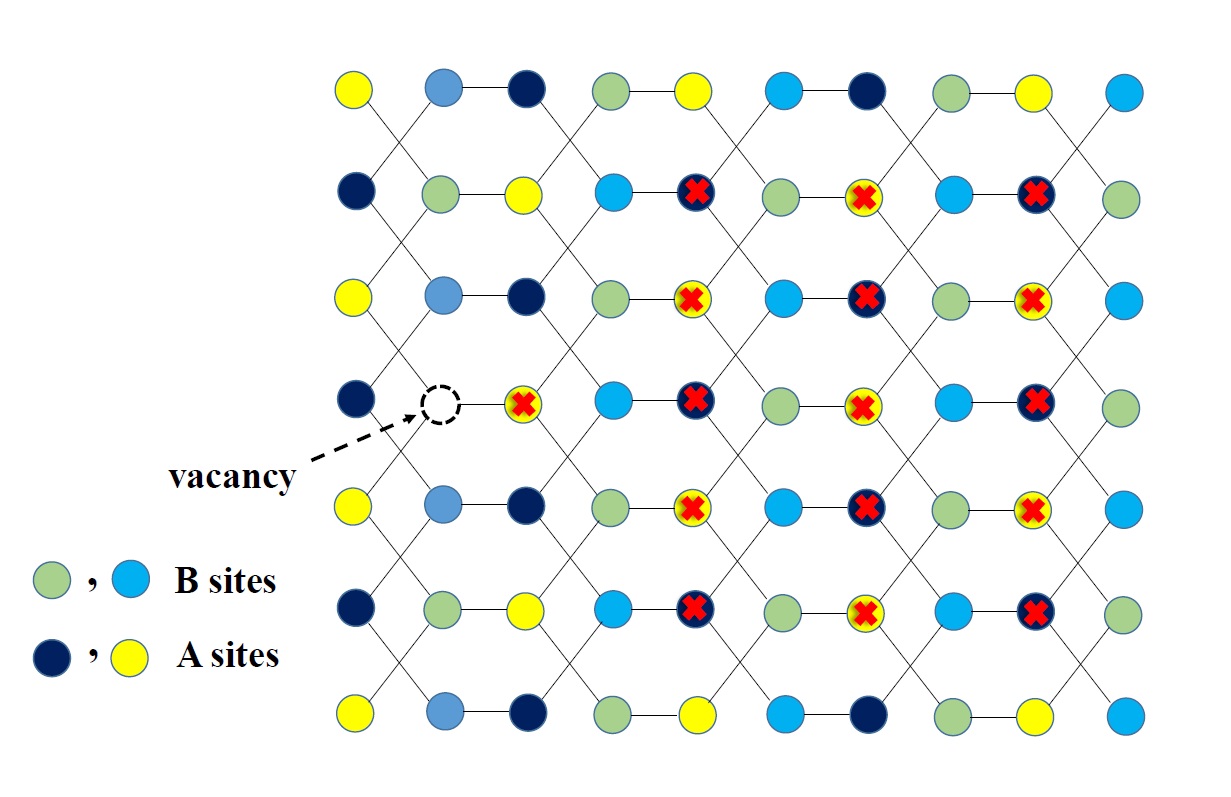}	
	\caption{\label{fig.vacancy}
		Armchair Bilayer Phosphorene Nanoribbon with the vacancy at B-type atom
	}
\end{figure}

Numerically, according to the Fig\ref{fig.vacancy}, the appearance of a single vacancy on the B-type atom will cause the localization to observe exactly on the A-type atoms around the vacancy which are marked with a times sign in the figure. Notwithstanding the foregoing, considering $\braket{A,i,j,k|\psi} =A_k(i,j)$, we have a set of equations:

\begin{flushleft}
	\begin{equation}
		\begin{split}
			\mathit{A}_1(0,0)t^{\parallel}_1+\mathit{A}_1(-1,1)t^{\parallel}_2 =0\\
			\mathit{A}_1(0,0)t^{\parallel}_1+\mathit{A}_1(0,1)t^{\parallel}_2 =0\\
			\mathit{A}_1(0,1)t^{\perp}_{11}+\mathit{A}_1(-1,1)t^{\perp}_{11}+\mathit{A}_2(0,1)t^{\parallel}_2 =0\\
			\mathit{A}_1(-1,1)t^{\perp}_{11}+\mathit{A}_2(-1,1)t^{\parallel}_2 =0\\
			\mathit{A}_1(0,1)t^{\perp}_{11}+\mathit{A}_2(1,1)t^{\parallel}_2 =0
		\end{split}
	\end{equation}
\end{flushleft}

which can be solved:
\begin{equation}
	\begin{split}
		\mathit{A}_1(-1,1) ,or \mathit{A}_1(0,1)&=-\mathit{A}_1(0,0)\frac{t^{\parallel}_1}{t^{\parallel}_2}\\
		\mathit{A}_2(1,1),or\mathit{A}_2(-1,1)&=\mathit{A}_1(0,0)\frac{t^{\parallel}_1t^{\perp}_{11}}{(t^{\parallel}_2)^2}\\
		\mathit{A}_2(0,1)&=2\mathit{A}_1(0,0)\frac{t^{\parallel}_1t^{\perp}_{11}}{(t^{\parallel}_2)^2}
	\end{split}
\end{equation}

\begin{figure}[h!]
	\includegraphics[height=0.90\linewidth]{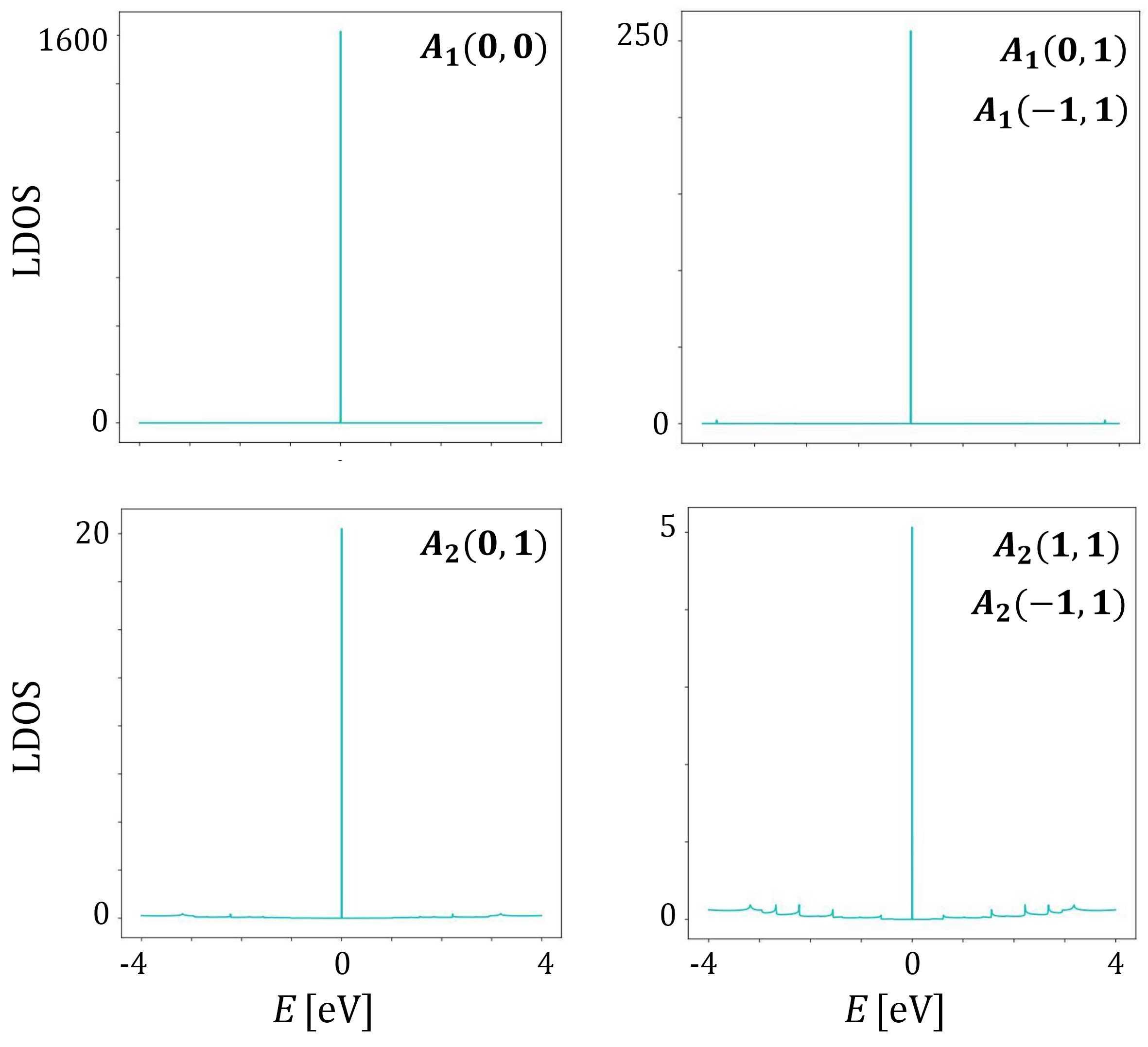}	
	\caption{\label{fig.LDOS}
		LDOS of nearest neighbors of a vacancy atom $(\mathit{B}_1(0,-1))$ in armchair Bilayer Phosphorene Nanoribbons.
	}
\end{figure}

Utilizing the result of a numerical method, the accuracy of the analytical results can be discussed and both results show that the LDOS values are symmetric about the center of the structure (Fig\ref{fig.LDOS}).

Also about next column:

\begin{equation}
	\begin{split}
		\mathit{A}_2(0,1)t^{\parallel}_1+\mathit{A}_2(-1,1)t^{\parallel}_1+\mathit{A}_2(-1,2) t^{\parallel}_2&=0\\
		\mathit{A}_2(-1,1)t^{\parallel}_1+\mathit{A}_2(-2,2)t^{\parallel}_2&=0\\
		(\mathit{A}_2(-1,2)+\mathit{A}_2(-2,2)) t^{\perp}_{11}+\mathit{A}_1(-1,2)t^{\parallel}_2
		+\mathit{A}_1(-1,1)t^{\parallel}_1&=0\\
		(\mathit{A}_1(-1,1)+x\mathit{A}_1(1,1))t^{\parallel}_1+\mathit{A}_1(0,2) t^{\parallel}_2\\
		+(\mathit{A}_2(0,2)+\mathit{A}_2(-1,2))t^{\perp}_{11}&=0
	\end{split}
\end{equation}

hence,

\begin{equation}
	\begin{split}
		\mathit{A}_2(0,2) or \mathit{A}_2(-1,2)&=-3\mathit{A}_1(0,0)\frac{(t^{\parallel}_1)^2 t^{\perp}_{11}}{(t^{\parallel}_1)^3}\\
		\mathit{A}_2(-2,2) or \mathit{A}_2(1,2)&=-\mathit{A}_1(0,0)\frac{(t^{\parallel}_1)^2 t^{\perp}_{11}}{(t^{\parallel}_1)^3}\\
		\mathit{A}_1(1,2) or \mathit{A}_1(-1,2)&=\mathit{A}_1(0,0)\frac{t^{\parallel}_1}{(t^{\parallel}_2)^2}(1+4\frac{t^{\parallel}_1( t^{\perp}_{11})^2}{(t^{\parallel}_2)^2})
	\end{split}
\end{equation}
In the presence of mentioned parameters, the results are exact and the analytical calculations are consistent with numerical results. Accordingly, the nearest site to the vacancy has a sharp peak of LDOS at zero energy. In contrast, when the rest of the hopping parameters are considered, the LDOS peak remains in the energy gap, but the change is that the LDOS peak shifts. So using numerical calculations, the energy variation obtains, and we can declare that form of the wave function in the presence of perturbation terms changes.
All of these studies are related to one vacancy, and it should be noted that considering two vacancies in one dimer(A and B sublattices), no impurity state is observed.

\section{Thermoelectric efficiency}

The thermoelectric power generation efficiency is calculated from the generated electrical power and the rate of heat flow, Eq.\ref{eq:cop}. Therefore, the generation of the highest electrical power in exchange for the lowest heat flow will be economically ideal.
\begin{equation}\label{eq:cop}
	\eta = \frac{P_{output}}{P_{input}}=\frac{VI}{\dot{Q}_h},
\end{equation}

In which $V$ is the applied bias voltage. Using the Landauer-B$\mathrm{\ddot{u}}$ttiker formalism, one can calculate the electric current. In the ballistic transport regime, the expression for calculating the current will then be
\begin{equation}\label{eq:current}
	I = \frac{2 e}{h} \int dE \; T_{LR}(E) \left( f_L-f_R\right),
\end{equation}

Here, $h$ and $e$ are the Planck constant and the elemental charge of one electron, respectively. $T(E)$ is the total transmission function. Also,   $f=1/\left[ e^{(E-E_F)/k_BT}+1\right]$ is the Fermi-Dirac distribution function and $k_B$ is Boltzmann constant.

The experission for calculating the electron contribution of heat current is the same as electric current but replacing the elemental charge $e$ with the $(E-\mu_h)$. 

\begin{equation}\label{eq:heatcurrent}
	\dot{Q}_h = \frac{2}{h} \int dE \; T_{LR}(E) (E - \mu_h)  \left( f_L-f_R\right).
\end{equation}

As stated in the introduction, increasing the thermoelectric efficiency of nanostructures is pursued in two approaches. The first is to optimize the electronic properties in order to increase the generated electric power and the second is to reduce the thermal conductivity in order to decrease the heat exchange in the system.  
In this study, we have focused on the shape of the transmission function to maximize the generated electric power and efficiency. 
According to the thermodynamics laws, the maximum efficiency (Carnot efficiency) of heat engines operating at two different temperatures is equal to
\begin{equation}\label{eq:Carnot}
	\eta_C=1-\frac{T_c}{T_h}
\end{equation} 
Carnot efficiency can only be achieved when the transmission function is equal to the delta function ($\delta(E-E_F)$) but such a transmission function leads to vanishing of the electric power output. 
The optimum efficiency at a given electric power is obtained for the case of boxcar shape transmission function ~\cite{Whitney14, Whitney15}. A boxcar transmission function acts as a band-pass filter and passes electrons in a special range of energy and blocks electrons out of this energy range. In the thermoelectric power generation process, the favorable electric current flows from hot lead to the cold side, and the current in the opposite direction is inappropriate. To enhance the electric power output and then thermoelectric efficiency, a boxcar transmission function only allowing ellectron flow in the desired direction while the destructive current in the opposite direction is eliminated.
The proper working temperature is determined by the width of the transmission window and should be in the order of $T\sim\frac{\Delta}{k_B}$, in which $T$ is the leads average temperature and $\Delta$ is width of the transmission window. The width of the transmission function determines the electrical output power and thermoelectric efficiency of the system. As $\Delta$ increases, the output power improves at the cost of decreasing efficiency and vise versa. Therefore, the optimal width depends on the required electrical output power and thermoelectric efficiency.

\section{Results}

	The system considered in this article is a bilayer armchair phosphorene nanoribbon that is connected to two hot and cold leads from both sides. The cold and hot lead temperatures are assumed to be $T_c = 150~\mathrm{K}$ and $T_h = 250~\mathrm{K}$, so the ideal Carnot efficiency at these temperatures is $40 \%$. In the center of the nanoribbon and in line with the armchair edge, we have periodically created vacancies. The distance between the vacancies is started from 4b (b is the lattice constant in armchair direction) and gradually increased. As seen in the previous sections, the presence of a vacancy leads to generation of a localized state, which is similar to a single energy level atom. The creation of periodic vacancies causes an overlap between the localized states and this set acts like a tight-binding chain.

\begin{figure}[h!]
	\includegraphics[height=0.8\linewidth]{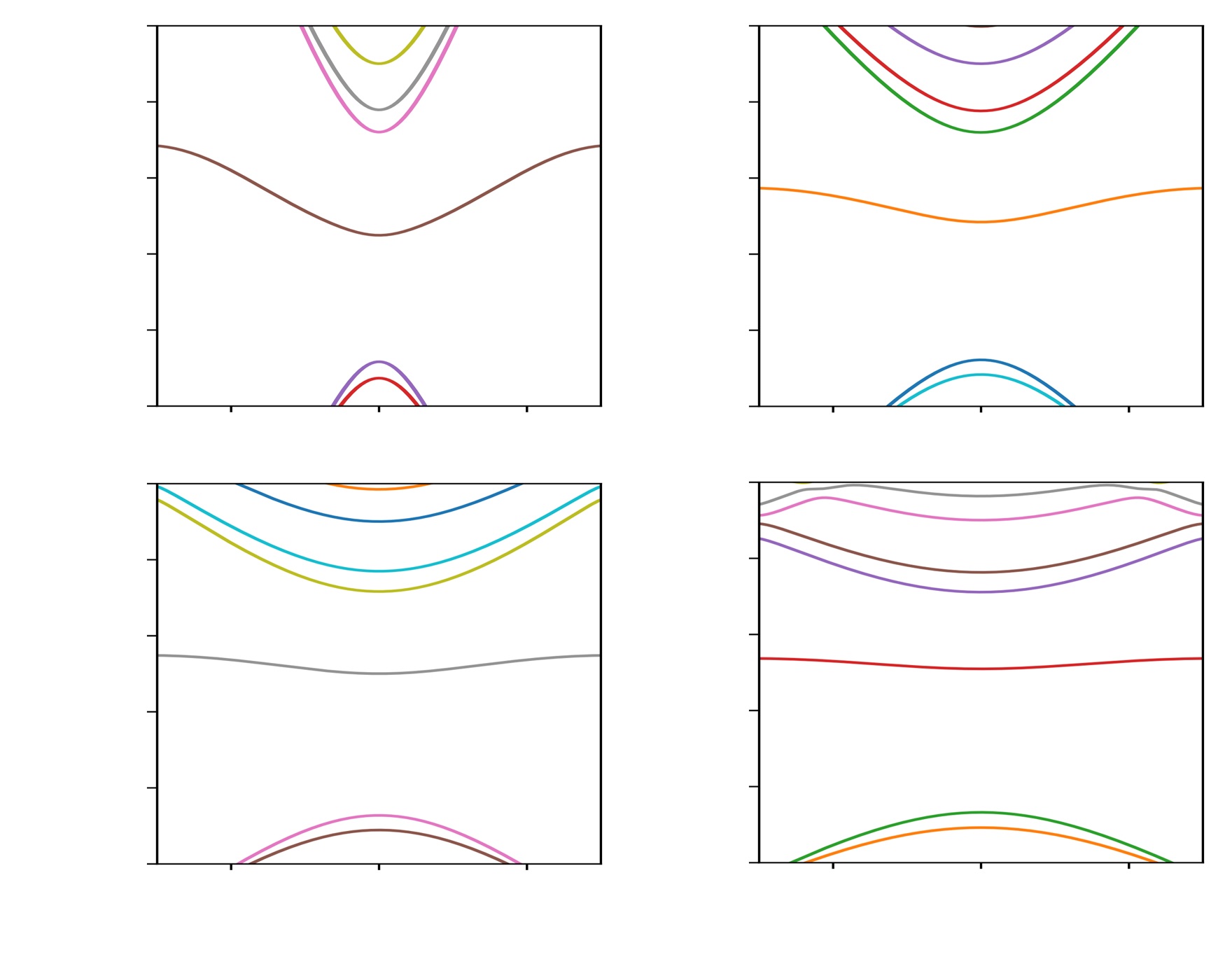}	
	\caption{\label{fig.bands} energy bandstructure of armchair bilayer phosphorene with periodic vacancies.
	}
\end{figure}

\begin{figure}[h!]
	\includegraphics[height=0.53\linewidth]{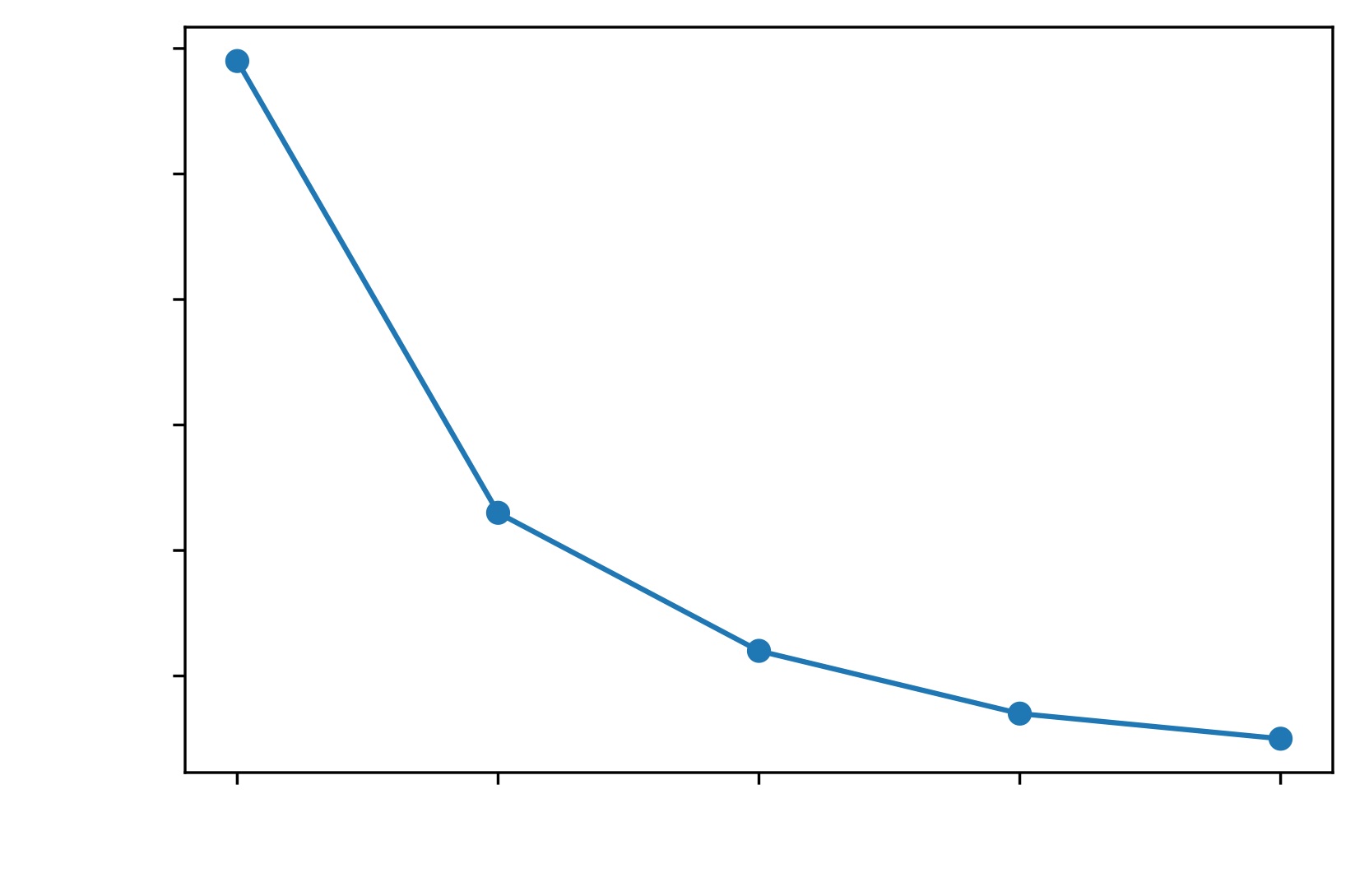}	
	\caption{\label{fig.width}
		the width of the impurity band transmission function as function of the distance between vacancies.
	}
\end{figure}

Figure \ref{fig.bands} demonstrates the energy bandstructure of considerd system for 4 different sapacing between vacancies. As can be seen, the periodic creation of vacancies has led to the emergence of a new band in the energy bandstructure, and the dispersion of this band decreases with the increase in the distance between the vacancies. This is because the overlap between the localized states decreases as the distance between the vacancies increases, and this leads to smaller hopping parameter in the chain of single energy level atoms. On the other hand, the energy band of such a chain is equal to $2t_{eff}\cos(k)$, Therefore, with the reduction of effective hopping parameter between the localized states of the periodic vacancies, the energy dispersion of this band decreases. The width of this energy band will determine the width of its transmission channel, and as can be seen in figure \ref{fig.width} one can tune the transmission channel by changing the spacing between vacancies. The width of this transmission channel is one of the most important factors that determine the thermoelectric properties of the system.

\begin{figure}[h!]
	\includegraphics[height=1.6\linewidth]{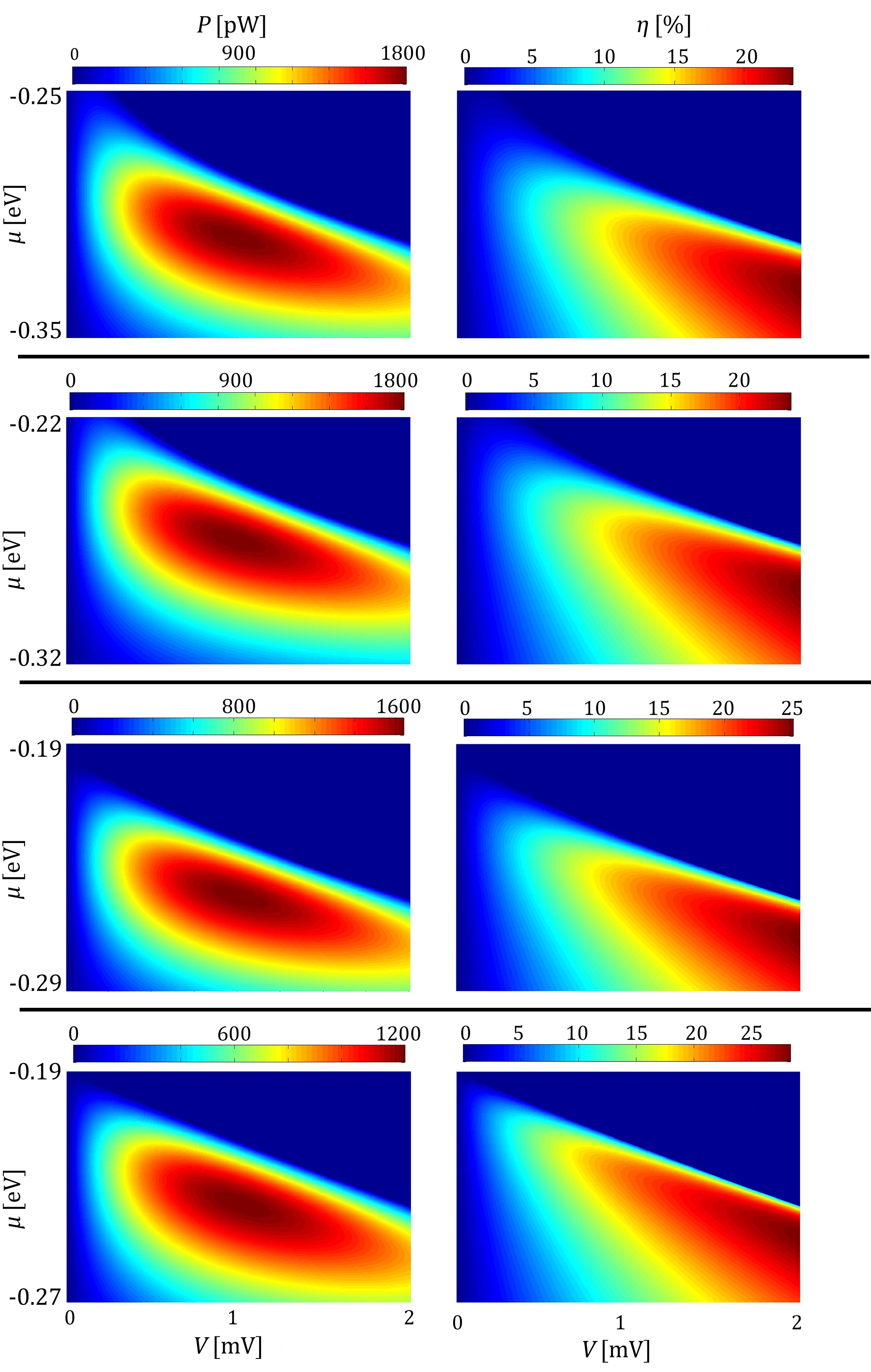}	
	\caption{\label{fig.Peff}
		electric power and thermoelectric efficiency for 4 values of spacing between vacancies.
	}
\end{figure}

Figures \ref{fig.Peff} show the electric power and thermoelectric efficiency, for 4 values of spacing between vacancies, plotted on color scale as functions of bias voltage $V$ and the average chemical potential $\mu$. 

We can attain optimal electric power and thermoelectric efficiency by properly regulating bias voltage and average chemical potential. 

As indicated in figure \ref{fig.Peff}, the highest maximum electric power is obtained for the case of $d=6b$, which has the widest transmission function $P^{d=6b}_{max}=1820~\mathrm{pW}$ while the thermoelectric efficiency at maximum electric power is larger for the case with larger spacing between vacancies which have a narrower transmission function $\eta^{d=12b}_{maxP}=20.3\%\simeq51\%\eta_C$.

\begin{figure}[h!]
	\includegraphics[height=0.85\linewidth]{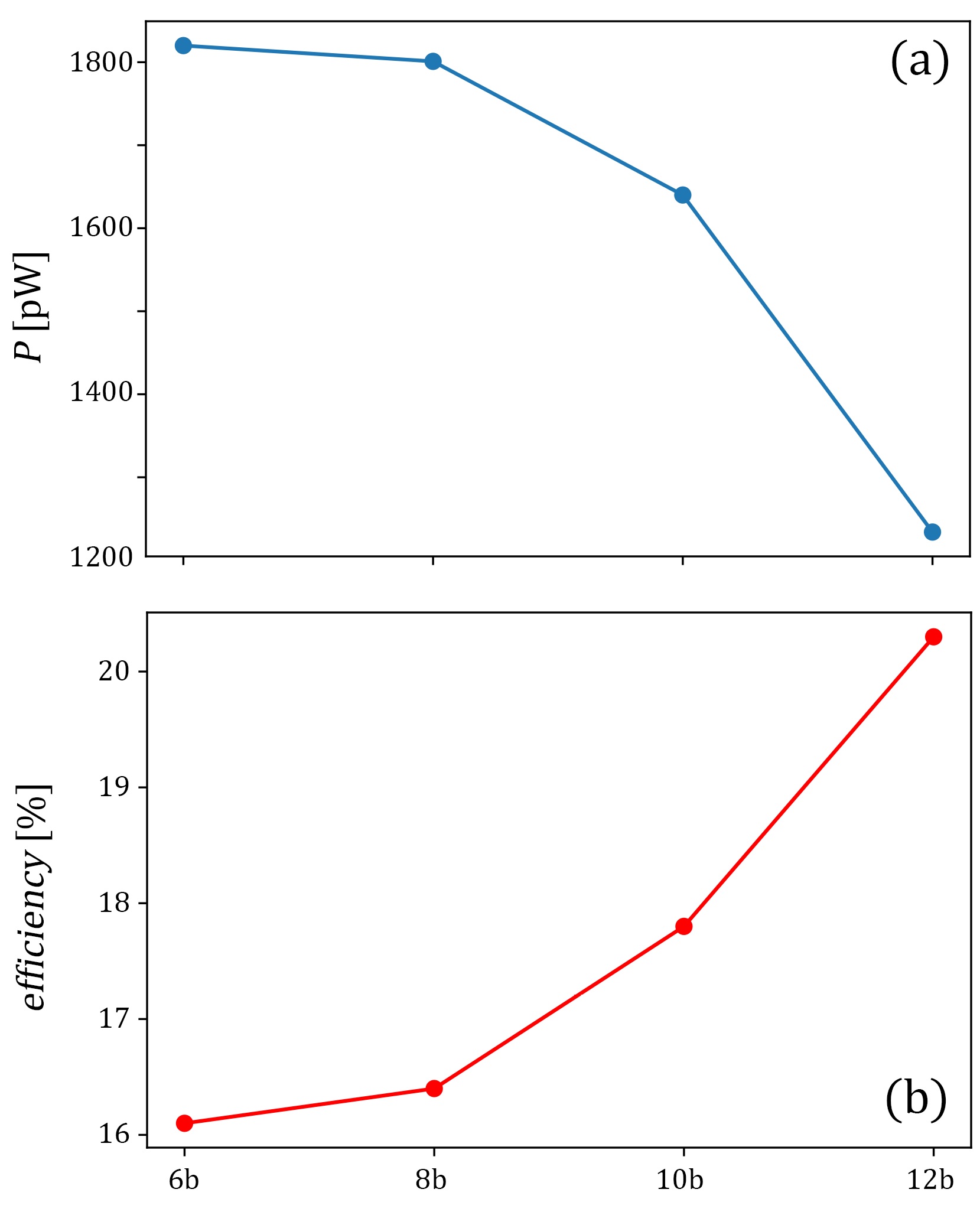}	
	\caption{\label{fig.Pwidth}
		variation of (a) electric power and (b) efficiency at maximum electric power as function of the distance between vacancies.
	}
\end{figure}

Figure \ref{fig.Pwidth}(a) shows the variation of electric power in terms of spacing between vacancies. As the distance between the vacancies increases, the electric power decreases. In the other cases $P^{d=8b}_{max}=1801~\mathrm{pW}$, $P^{d=10b}_{max}=1640~\mathrm{pW}$, and $P^{d=12b}_{max}=1234~\mathrm{pW}$.

Figure \ref{fig.Pwidth}(a) illustrates that in contrast to maximum electric power, the efficiency at maximum electric power increases by increasing the spacing between vacancies, $\eta^{d=6b}_{maxP}=162.1\%$, $\eta^{d=8b}_{maxP}=16.04\%$, and $\eta^{d=10b}_{maxP}=17.8\%$.

It should be mentioned that choosing the distance between vacancies basically depends on two points. The first point is the temperatures of two leads, and for the higher temperatures, we need to choose cases with a wider transmission channel. The second point is how much electrical power and thermoelectric efficiency is required.

\section{Conclusion }
\label{sec.conclu}

	In conclusion, we have studied the thermoelectric power generation properties of armchair bilayer phosphorene nanoribbons. we have shown that creation of periodic vacancies at the middle of nanoribbon in armchair direction leads to generation of localized states. The overlap of such states will lead to the appearance of a new band in the energy bandstructure. 
In the following, we have shown that the change of the distance between the vacancies leads to the change of the transmission cannel width of these states and, as a result, the electric power and thermoelectric efficiency of the system.

\bibliographystyle{elsarticle-num}
\bibliography{cite}

\end{document}